\documentclass[aps,twocolumn,prl,preprintnumbers,superscriptaddress]{revtex4-1}
\pdfoutput=1
\usepackage{amsmath,amssymb,bm,comment}
\usepackage{graphicx}
\usepackage{hyperref}

\begin{document}

\preprint{
}

\title{Anomalous hydrodynamics kicks neutron stars}
\author{Matthias Kaminski}
\affiliation{Department of Physics, University of Washington, Seattle, WA 98195, USA}
\affiliation{Department of Physics and Astronomy, University of Alabama, Tuscaloosa, AL 35487, USA}
\affiliation{Department of Physics and Astronomy, University of Victoria, Victoria, BC, V8P 5C2, Canada}
\author{Christoph F.~Uhlemann}
\affiliation{Department of Physics, University of Washington, Seattle, WA 98195, USA}
\author{Marcus Bleicher}
\affiliation{Frankfurt Institute for Advanced Studies, Goethe-Universit\"at Frankfurt, Germany}
\affiliation{Institut f\"ur Theoretische Physik, Goethe Universit\"at Frankfurt, Germany}
\author{J\"urgen Schaffner-Bielich}
\affiliation{Institut f\"ur Theoretische Physik, Goethe Universit\"at Frankfurt, Germany}
\date{\today}

\begin{abstract}
Observations show that, at the beginning of their existence, neutron stars are accelerated briskly to velocities of up to $1000$ km/s.
We discuss possible mechanisms contributing to these kicks in a systematic effective-field-theory framework.
Anomalies of the underlying microscopic theory result in chiral transport terms in the hydrodynamic description,
and we identify these as explanation for the drastic acceleration.
In the presence of vorticity or a magnetic field, the chiral transport effects cause neutrino emission along the respective axes.
In typical scenarios, the transport effect due to the magnetic field turns out to be strong enough to explain the kicks.
Mixed gauge-gravitational anomalies enter in a distinct way, and we also discuss their implications.
\end{abstract}

\maketitle

\section{Introduction}
In this letter we study transport effects in neutron stars in an effective-field-theory framework, utilizing hydrodynamics.
Proto-neutron stars are observed to receive kicks, i.e.\ a large change of momentum along their axis of rotation early in their evolution~\cite{Chatterjee:2005mj}.
On a qualitative level, these kicks have been linked to asymmetric neutrino emission already in \cite{Vilenkin:1979ui, Vilenkin:1978is, PhysRevD.22.3080}.
The modern formulation of hydrodynamics allows us to give a systematic and quantitative analysis.
The crucial ingredient are quantum effects, 
which often only enter as small corrections to classical computations and rarely show up on macroscopic scales.
They do make a qualitative difference in theories with anomalies, though: when classical conservation laws are broken by quantum effects.
Famously, this explains the decay of the pion~\cite{Adler:1969gk,Bell:1969ts}.
The presence of anomalies in a microscopic theory is a robust feature
that persists in effective-field-theory descriptions~\cite{'tHooft:1979bh}.
More recently, anomalies were found to have striking implications in the hydrodynamic regime \cite{Erdmenger:2008rm, Banerjee:2008th, Son:2009tf}.
The effects of the resulting new transport phenomena have mostly been studied on microscopic 
length scales, in the context of heavy-ion-collisions \cite{Kharzeev:2010gr, Kharzeev:2004ey, Fukushima:2008xe}.
In this letter, we show that anomalous transport in hydrodynamics can have sizable effects also on 
very macroscopic scales, and explain the neutron star kicks.

In the first part of this letter we discuss the hydrodynamic framework and anomalous transport effects
on a general level, with emphasis on the features that will be relevant for our effective description of a neutron star.
In the second part we discuss the specific currents which receive large contributions from anomalous
transport effects in a typical neutron star, and -- as we argue -- can propel it to the observed velocities.
The resulting mechanism is illustrated in Fig.~\ref{fig:emission}.

\section{Anomalous hydrodynamics}\label{sec:anomalous-hydro}

\begin{figure}
\includegraphics[width=0.2\textwidth]{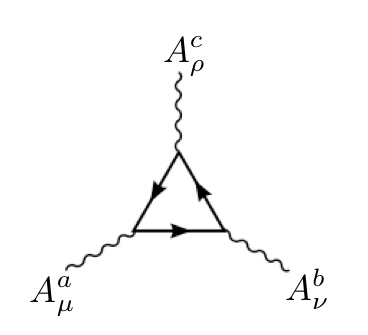}
\caption{\label{fig:triangleDiagram}
Anomalies appear as violation of the Ward identities for triangle diagrams of the form shown above, 
giving a convenient way to compute the coefficients $C_{abc}$.}
\end{figure}

In recent years, hydrodynamics has been reinterpreted and developed systematically in 
effective field theory language~\cite{Baier:2007ix, Bhattacharyya:2008jc}. 
One striking result of this program is that anomalies of the underlying microscopic quantum field theory 
cause macroscopic transport effects~\cite{Erdmenger:2008rm,Banerjee:2008th,Son:2009tf}. 
Any system which is described microscopically by a relativistic quantum field theory with anomalies 
receives the following contributions to a current corresponding to a global symmetry
(at first order in the hydrodynamic expansion in gradients)~\cite{Son:2009tf}
\begin{equation}\label{eq:anomalousCurrent}
J_a^\mu = n_a u^\mu + \sigma_a^{\hphantom{a}b} V_{b}^\mu +  \sigma_a^V \omega^\mu + \sigma_{ab}^B B^{b\,\mu} +\mathcal{O}(\partial^2) \, ,
\end{equation}
where $a,b$ label the currents in the theory, $n_a$ is the net charge density,
$u^\mu$ is the fluid velocity, $\sigma_a^{\hphantom{a}b}$ is the conductivity
and $V_a^\mu=(E_a^\mu - T (\eta^{\mu\nu} +u^\mu u^\nu) \partial_\nu \frac{\mu_a}{T})$ with the field strength $E_a$. 
Furthermore, we have the temperature $T$, flat metric $\eta^{\mu\nu}$ and the chemical potential $\mu_a$ which is thermodynamically 
conjugate to $n_a$. The two remaining terms in equation~\eqref{eq:anomalousCurrent} contain the chiral vortical coefficient $\sigma^V_a$, 
the vorticity $\omega^\mu=\frac{1}{2}\epsilon^{\mu\nu\rho\sigma} u_\nu \partial_\rho u_\sigma$, the chiral magnetic coefficient $\sigma_{ab}^B$,
and $B_b^\mu=\frac{1}{2}\epsilon^{\mu\nu\rho\sigma} u_\nu F_{b\,\mu\nu}$ with the field strength tensor $F_b$ 
(for the electromagnetic U(1)$_\mathrm{em}$ this is the familiar magnetic field).
The remarkable feature is that,
using the standard hydrodynamic restriction of positivity of the local entropy production,
the
transport coefficients $\sigma_a^V$ and $\sigma_{ab}^B$ can be computed exactly from the 
anomalies of the underlying theory and thermodynamic quantities~\cite{Son:2009tf,Neiman:2010zi}.
The explicit expressions depend on the chosen frame.
The commonly used Landau frame fixes the fluid velocity through conditions on momentum transport,
which makes chiral transport effects partly implicit. 
This is inconvenient for a study of their implications, and we therefore use the fixed 
laboratory frame discussed in \cite{Landsteiner:2012kd}, where 
\begin{align} \label{eqn:xi}
\sigma_a^V &= \frac{1}{2}C_{abc} \mu^b \mu^c-\beta_aT^2 \,, 
&
\sigma_{ab}^B &= C_{abc} \mu^c  \,.
\end{align}
The $C_{abc}$ are the coefficients characterizing the anomalous conservation laws,
$\langle\partial_\mu J_a^\mu\rangle = \frac{1}{8} C_{abc} \epsilon^{\mu\nu\rho\sigma} F^b_{\mu\nu} F^c_{\rho\sigma}$. 
In perturbative calculations anomalies arise from triangle diagrams of the form shown in Fig.~\ref{fig:triangleDiagram}, 
involving the three currents $j_{a/b/c}$. 
The diagrams are generally not anomalous when all currents are vector-like~(V), 
but can be for diagrams with axial-vector (A) contributions of the form VVA or AAA.
The $T^2$-term in \eqref{eqn:xi} encodes the chiral transport effect at zero chemical potential.
Remarkably, the coefficients $\beta_a$ can be linked to mixed gauge-gravitational anomalies of the form
$\nabla_\mu J^\mu_{\text{cov}} = \frac{1}{4} \epsilon^{\mu\nu\rho\sigma} c_m R^\alpha_{\beta\mu\nu} R^\beta_{\alpha\rho\sigma} +\dots$
\cite{Landsteiner:2011cp}, where $\nabla$ is the covariant derivative, 
$R$ the Riemann curvature and $c_m$ a numerical coefficient.
The analysis of \cite{Jensen:2012kj} showed that $\beta = -8\pi^2 c_m$ for a theory with a single global U(1) symmetry.
Just as no background gauge fields are needed in order to see effects of chiral anomalies in the hydrodynamic description,
no curved space or large gravitational fields are needed to get a sizable contribution from the mixed
gauge-gravitational anomalies.
When dynamical gauge fields contribute to the anomalies, the transport coefficients are not protected from 
renormalization and may receive additional contributions~\cite{Golkar:2012kb}.

The explicit expressions for $\sigma^V$, $\sigma^B$ in \eqref{eqn:xi} will allow us to compute the 
magnitude and direction of chiral transport contributions in a proto-neutron star below.

\section{Currents and Anomalies}

In this section we introduce the relevant currents and discuss their anomalies.
For the neutron star kicks we will be interested in the leptonic currents,
and to keep the discussion clear we make a number of simplifying assumptions:
We ignore neutrino masses, since they are very small compared to typical temperatures and chemical potentials
in a neutron star, and we will not speculate about the existence or nature of right-handed neutrinos.
This means we do not take into account neutrino oscillations and the various lepton flavors are classically conserved separately.
The relevant leptons for our purposes are electrons and electron neutrinos, 
since these are the flavors mostly produced in the relevant electroweak processes \cite{Prakash:1996xs,Pons:1998mm}.
For our purposes lepton number therefore means electron number.
The electron and neutrino currents alone are not conserved at the classical level due to the weak interactions, 
and for the hydrodynamic description we therefore consider the classically conserved lepton number current combining both.
At typical neutron star temperatures of $\mathcal O(10\,\mathrm{MeV})$, sphaleron processes are suppressed \cite{Kuzmin:1985mm} 
and we will not take them into account.
The electron mass is small compared to typical temperatures and chemical potentials as well, 
and it is tempting to just work with massless electrons.
Despite being small, the electron mass was found to have drastic implications for the asymmetry between left-handed 
and right-handed electrons generated during the formation of a neutron star in \cite{Grabowska:2014efa}.
We will assume here that the left-handed and right-handed lepton number currents, $J_{\ell L}$ 
and $J_{\ell R}$ are separately conserved within each local equilibration region to a good enough accuracy 
to be part of the hydrodynamic description.
The holographic study in \cite{Jimenez-Alba:2014iia} has shown that the anomalous transport effects present for 
conserved currents persist when the conservation is slightly violated, and they were even enhanced in certain cases.
We leave the question of whether or not a large electron asymmetry is generated during the formation of the neutron
star open and consider both scenarios when we discuss anomalous transport in the next section.

To discuss the anomalies we will use the linear combinations $J_{\ell L}\pm J_{\ell R}$ and call them $J_{\ell}$ and $J_{\ell 5}$,
respectively.
The electron part in these currents is vector/axial vector like, while the neutrino part is purely left-handed for both.
The charges under the respective symmetries U(1)$_{\ell/\ell5}$ are $1$ for all fields except for the right-handed electrons,
which have $-1$ w.r.t.\ U(1)$_{\ell 5}$. 
Since neither of $J_{\ell}$ and $J_{\ell 5}$ are purely vector like, we get a rather large number of different 
anomalous triangle diagrams.
To begin with, both symmetries have a U(1)$_a^3$ anomaly, yielding non-vanishing coefficients $C_{aaa}$ with $a=\ell,\ell 5$.
We also get non-vanishing $C_{\ell,\ell,\ell 5}$ and $C_{\ell,\ell 5,\ell 5}$.
These will be relevant for the chiral effects due to the vorticity only.
We also have mixed anomalies with the electromagnetic gauge field.
Since the U(1)$_\mathrm{em}$ is vector like, we get these from VVA diagrams.
In these diagrams the neutrinos do not contribute since they are not charged under U(1)$_\mathrm{em}$,
and $J_{\ell/\ell5}$ therefore actually behave vector/axial vector like.
We get two non-vanishing anomaly contributions
corresponding to $C_{\mathrm{em},\ell,\ell 5}$ and $C_{\mathrm{em},\mathrm{em},\ell 5}$.
Computing the actual numerical values of all these coefficients is straightforward, and we do not
need to list them here.

\section{Chiral transport in proto-neutron stars}
After the discussion of the general framework and the relevant currents above,
we now focus on anomalous transport of leptons in the bulk of a neutron star.
Electrons and neutrinos appear together in the classically conserved currents $J_{\ell}$ and $J_{\ell5}$,
and anomalous transport, if present, thus affects both.
The transparency properties of the crust will be discussed below.

To estimate the relative strength of the two anomalous transport effects in the neutron star we take a look at its vorticity.
The star can be modeled as a rigidly rotating disk of radius $r_\mathrm{N}$,
with vorticity $\omega=-2 \Omega$ 
where $\Omega$ is the angular velocity.
With $\Omega=2\pi/\mathrm{ms}$ as a ballpark figure \cite{Ott:2005wh}, we then find $\omega\approx10^{-17}\,\mathrm{MeV}$.
The magnetic fields, on the other hand, can easily take values of $10^{12}\,\mathrm{G}\approx 0.1\,\mathrm{MeV}^2$ \cite{2013arXiv1305.2542R}.
The dimensionful quantities entering the coefficients \eqref{eqn:xi} are all of $\mathcal O(\mathrm{MeV})$, and we thus expect the chiral effects due to the magnetic field to 
be dominant by many orders of magnitude.

For an order-of-magnitude estimate of the coefficient for the 
chiral effect due to vorticity we use \cite{Prakash:1996xs,Pons:1998mm}
\begin{align}\label{eqn:mu-epsilon}
\epsilon/n_\ell&=\mu^\ell=300\,\mathrm{MeV}~,& T&=10\,\mathrm{MeV}~,
\end{align}
along with $\epsilon=3P$. This yields
\begin{align} 
 \sigma_{\ell}^V\approx (10^3C_x-10^2\beta)\,\mathrm{MeV}^2~,
\end{align}
with an $\mathcal O(1)$ coefficient $C_x$ parametrizing the contribution from U(1) anomalies
and the second term representing the temperature-dependent contributions.
We see that the coefficient $\beta$, which includes the gravitational contributions, 
enters at essentially the same order of magnitude as the pure U(1) anomalies parametrized by $C_x$.

We now turn to the chiral effects due to the magnetic field, which we discuss in more detail.
The coefficients we are interested in are $\sigma_{a,\mathrm{em}}^B$ with $a=\ell,\ell5$\,, 
such that $B$ is the magnetic field of U(1)$_\mathrm{em}$. 
The explicit form is
$\sigma_{a,\mathrm{em}}^B = C_{a,\mathrm{em},c} \mu^c$.
We see that at least one of the external fields in the triangle diagrams computing the $C_{abc}$ is the 
electromagnetic gauge field.
As explained above, in that case the only non-vanishing coefficients are $C_{\mathrm{em},\ell,\ell 5}$ and $C_{\mathrm{em},\mathrm{em},\ell 5}$.
Assuming that the neutron star is neutral to a good approximation, we ignore the contribution due to the latter.
With $C\,{=}\,C_{\mathrm{em},\ell,\ell5}\,{=}\,1/(2\pi^2)$ the explicit form of the coefficients becomes
\begin{align} \label{eq:xiLeptonEM}
\sigma_{\ell,\mathrm{em}}^B &= C\mu^{\ell 5}\,,& \sigma_{\ell5,\mathrm{em}}^B &= C\mu^{\ell}\,.
\end{align}

To estimate the resulting currents we use \eqref{eqn:mu-epsilon} for the values of $\mu^\ell$ and $n_\ell$,
and for the corresponding values for $J_{\ell 5}$ we discuss two cases.
The first case is $\mu^{\ell 5}\,{\approx}\,0$.
Noting that $n_{\ell 5}\,{=}\, n_{\mathrm{e_L}}\,{-}\, n_{\mathrm{e_R}}\,{+}\,n_{\mathrm{\nu}}$,
this describes the case where electron chirality is preserved:
the electroweak interactions may generate a large chiral asymmetry for electrons,
but the combined number of left-handed electrons and neutrinos is conserved.
The range of magnetic fields observed in neutron stars is rather wide, and spans several orders of magnitude \cite{2013arXiv1305.2542R}.
With the intermediate value $B=0.1\,\mathrm{MeV}^2$ and (\ref{eqn:mu-epsilon}) we find
\begin{align}\label{eqn:J-B-estimate}
 \vec{J}_\ell&\approx 0~,&
 \vec{J}_{\ell5}=C\mu^\ell\vec{B}\approx  \vec{e}_B\cdot1\,\mathrm{MeV}^3~.
\end{align}
The effect  is illustrated in Fig.~\ref{fig:emission}(b):
$J_\ell\,{\approx}\, 0$ means that there is an equal number of leptons moving parallel and antiparallel to the magnetic field.
From the non-vanishing $J_{\ell 5}$ we conclude that left-handed and right-handed leptons are on average moving in opposite 
directions.

We now come to the second scenario, where the chiral asymmetry of the electrons is washed out during the formation of the 
neutron star \cite{Grabowska:2014efa}.
In that case we do get a non-vanishing $\mu^{\ell5}$ from the excess of left-handed particles due to the neutrinos.
The number of left-handed and right-handed leptons in Fig.~\ref{fig:emission}(b) then is not equal, 
resulting in a non-vanishing $J_\ell$.
The left-handed current $(J_\ell+J_{\ell 5})/2$, however, changes only by an $\mathcal O(1)$ factor compared to the previous scenario,
and the same applies for its composition in terms of electrons and neutrinos.
The crucial point for us is that only the neutrinos will be able to leave the neutron star, 
and the number of excess neutrinos moving along the magnetic field changes only by an $\mathcal O(1)$ factor.
The order-of-magnitude estimate of the kick in the next section is therefore not affected.

\section{Kicks from chiral transport in proto-neutron stars}
\begin{figure}[h]
\includegraphics[width=0.4\textwidth]{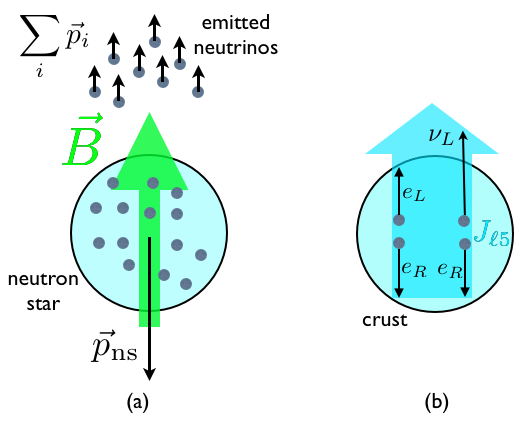}
\caption{\label{fig:emission} 
{\it A neutron star rocket with neutrino propulsion:} 
Neutrinos are emitted from a proto-neutron star through chiral transport effects parallel to the magnetic field $\vec B$. 
(a) The general mechanism: each neutrino carries away the momentum $\vec p_i$, 
producing a recoil $\vec p_{\text{ns}} = -\sum_i \vec p_i$ on the neutron star.
(b) Illustration of the currents: $J_\ell\,{\approx}\,0$ but sizable $J_{\ell 5}$ means that
left-handed leptons flow opposite to right-handed ones. 
Only the left-handed neutrinos can escape through the crust.
}
\end{figure}

With the precise form of the transport coefficients and an estimate for the resulting currents, 
we can now estimate whether and how efficiently the resulting currents can accelerate the neutron star. 
In typical scenarios, the crust of a neutron star is transparent only to neutrinos,
which are thus the only particles emitted, as illustrated in Fig.~\ref{fig:emission}.
This produces a recoil on the neutron star, which we estimate as follows.

To get the number of neutrinos leaving the neutron star, we compute the neutrino flux
$\dot N_\nu=|\vec{J}|A$, with the area of the rotating disk $A=\pi r_\mathrm{N}^2$ and 
$r_\mathrm{N}=10\,\mathrm{km}$.
For the current we take the value for $J_{\ell 5}$ given in (\ref{eqn:J-B-estimate}), 
augmented by a fudge factor $1/2$ to account for the fact that only the neutrinos can leave the neutron star.
This is compatible with the number densities given in \cite{Prakash:1996xs,Pons:1998mm}, 
but should be understood as order-of-magnitude estimate.
Converting to SI units, we find $\dot N_\nu\,{\approx}\, 10^{54}/\mathrm{s}$.
To compute the corresponding momentum current we could use the chiral transport coefficients for the 
energy-momentum tensor, as given in \cite{Landsteiner:2012kd}.
For the sake of simplicity, however, we just use the Fermi momentum as average momentum per 
neutrino, $\langle p_\nu\rangle\,{\approx}\, \mu^\ell$, which reproduces the same result.
For the momentum of the neutron star after the kick we then get
$\Delta P_\mathrm{NS}\,{=}\,\Delta t \dot N_\nu \langle p_\nu\rangle$,
where $\Delta t\,{\approx}\, 10\,\mathrm{s}$ is the time span we assume for the kick to last.
With a neutron star mass of $m_\mathrm{NS}\,{=}\,3\cdot 10^{30}\,\mathrm{kg}$ this yields
\begin{align}
 \Delta v&\approx 10^3\,\mathrm{km/s}~.
\end{align}
We thus find that the sudden momentum gains can indeed be explained by rapid neutrino emission 
due to the chiral transport effects, resulting in the simple picture shown in Fig.~\ref{fig:emission}.

\section{Discussion}
In this letter we have estimated anomalous transport effects in proto-neutron stars in a systematic hydrodynamic framework.
There are two independent effects, one causing neutrino emission along the axis of rotation, and the other one causing emission along the magnetic field of the proto-neutron star.
The latter turns out to be dominant by many orders of magnitude, and the neutrino recoil can indeed accelerate a typical proto-neutron star 
to velocities of order $10^3\,\mathrm{km/s}$, in agreement with observations. 
The specific numbers given in the text are based on order-of-magnitude estimates for the properties of a neutron star, 
but we have provided the framework and formulae to perform accurate calculations.

At the early times when the neutron star kicks happen, the crust is typically transparent for neutrinos only.
Studies of anomalous transport effects for electrons alone \cite{Charbonneau:1,Charbonneau:2} could therefore not explain the early kicks.
In the hydrodynamic framework we had to consider the classically conserved lepton number currents involving electrons and neutrinos. 
The electrons are crucial in the bulk of the neutron star and are only filtered out at the crust.
This leaves the neutrinos to escape and kick the neutron star.
Another difference to previous approaches is that we work in an effective-field-theory description from the outset.
In previous studies of asymmetric neutrino emission as kick mechanism \cite{Sagert:2007as,Vilenkin:1979ui, Vilenkin:1978is, PhysRevD.22.3080},
an asymmetry produced by processes studied at the microscopic level had to be evolved to macroscopic scales, and there suffered from thermal wash-out.
Our mechanism starts out directly with a macroscopic parity-violating transport effect on the level of the effective 
hydrodynamic description. A short mean free path here is a necessary ingredient for the hydrodynamic description to be valid,
and not a problem.

The precise form of the transport terms also allows for phenomenological conclusions.
On a qualitative level, we expect the kick to be aligned with the axis of rotation only if the 
magnetic field is aligned with it.
More quantitatively, we find a precise relation between the properties of the neutron star and the strength and direction of the kick.
The chiral effect due to the vorticity in principle offers access to mixed gravitational anomalies, 
which result in a quadratic temperature dependence.
For typical neutron stars the effect is outshined by the chiral effects due to the magnetic field, 
but there may be situations where this is different.
Finally, we note that the transport coefficients could also be sensitive to torsional contributions to the anomalies,
discussed recently in \cite{Parrikar:2014usa}, but leave a detailed analysis for the future.

{\it{Acknowledgements.}} We are grateful to K.~Jensen, A.~Karch, K.~Landsteiner, A.~Nelson, S.~Reddy and L.~Yaffe for very helpful discussions.
We also thank S.~Reddy for sharing \cite{Grabowska:2014efa} prior to publication.
This work has been supported in part by the US Department of Energy under contract number DE-SC0011637.
CFU is supported by a DFG research fellowship.

\bibliographystyle{h-physrev}
\bibliography{kicks}
\end{document}